\documentclass[12pt]{article}
\usepackage{amscd,amssymb,amsmath,latexsym}
\usepackage[mathscr]{euscript}
\usepackage{epsfig}
\usepackage{fancybox}
\title{Smoothness of correlations in the Anderson model\\ 
at strong disorder}

\author{Jean V. Bellissard$^{1}$\footnote{Partially supported by NSF Grant
0300398.},
Peter D.\ Hislop$^{2}$\footnote{Partially supported by NSF Grant 0503784}\\
{\small $^1$ Georgia Institute of Technology, School of Mathematics,
Atlanta, GA 30332-0160}\\
{\small $^2$ Department of Mathematics, University of Kentucky,
Lexington, KY  40506-0027}}

\newtheorem{theo}{Theorem}
\newtheorem{defini}{Definition}
\newtheorem{proposi}{Proposition}
\newtheorem{lemma}{Lemma}
\newtheorem{coro}{Corollary}
\newtheorem{rem}{Remark}

\newtheorem{prob}{Problem}

\newcommand{\Bb}{{\mathcal B}}
\newcommand{\Cc}{{\mathcal C}}
\newcommand{\Dd}{{\mathcal D}}
\newcommand{\Ee}{{\mathcal E}}

\newcommand{\Hh}{{\mathcal H}}

\newcommand{\Nn}{{\mathcal N}}

\newcommand{\Tt}{{\mathcal T}}
\newcommand{\Uu}{{\mathcal U}}
\newcommand{\Vv}{{\mathcal V}}

\newcommand{\id}{{\mathbf 1}}

\newcommand{\CM}{{\mathbb C}}
\newcommand{\EM}{{\mathbb E}}

\newcommand{\NM}{{\mathbb N}}

\newcommand{\PM}{{\mathbb P}}
\newcommand{\RM}{{\mathbb R}}

\newcommand{\ZM}{{\mathbb Z}}

\newcommand{\HG}{{\mathfrak H}}

\newcommand{\JV}{\vec{J}}

\newcommand{\RV}{\vec{R}}

\newcommand{\naV}{\vec{\nabla}}

\newcommand{\bs}{{\mathscr B}}

\newcommand{\ps}{{\mathscr P}}

\newcommand{\fU}{{\underline{f}}}
\newcommand{\lU}{{\underline{l}}}
\newcommand{\mU}{{\underline{m}}}
\newcommand{\nU}{{\underline{n}}}
\newcommand{\sU}{{\underline{s}}}
\newcommand{\tU}{{\underline{t}}}

\newcommand{\zU}{{\underline{z}}}
\newcommand{\AU}{{\underline{A}}}
\newcommand{\EU}{{\underline{E}}}
\newcommand{\OU}{{\underline{0}}}
\newcommand{\epsU}{{\underline{\epsilon}}}
\newcommand{\sigU}{{\underline{\sigma}}}

\newcommand{\TR}{{\rm Tr}}                         
\newcommand{\TV}{\Tt_{\PM}}                        
\newcommand{\Cs}{$C^{\ast}$-algebra }              
\newcommand{\tra}{\mbox{\sc t}}                    
\newcommand{\trb}{\mbox{\footnotesize\sc t}}       
\def\essup{\mbox{\rm $\PM$-essup}}                 
\def\conv{\mbox{\rm Conv}}                         
\def\Res{\mbox{\rm Res}}                           

\begin{document}

\maketitle

\begin{abstract}
We study the higher-order correlation functions of covariant families of
observables associated with
random Schr\"odinger operators on the lattice in the strong disorder regime.
We prove that if the
distribution of the random variables has a density analytic in a strip
about the real axis, then these correlation
functions are analytic functions of the energy
outside of the planes corresponding to
coincident energies. In particular, this implies
the analyticity of the density of states,
and of the current-current correlation function outside of
the diagonal. Consequently, 
this proves that the current-current correlation function
has an analytic density 
outside of the diagonal at strong disorder.
\end{abstract}


\tableofcontents

\vspace{.5cm}

\noindent KEY WORDS: {\em random operators, transport, correlations}

\vspace{.3cm}

\section{Correlation Functions}
\label{highcorr05.sec-intro}

\noindent The study of the higher-order correlation functions for random
Schr\"odinger operators is essential for an understanding of the transport
properties of the system. In this paper, we study the correlation functions
for covariant families of observables associated with random Anderson-type
Schr\"odinger operators on $d$-dimensional lattices $\ZM^d$ in the strong
disorder regime. The Anderson model is given by the following random
Hamiltonian acting on $\ell^2(\ZM^d)$

\begin{equation}
\label{highcorr05.eq-andersonmodel}
H_{\omega}\psi(x) \;=\; \lambda
\sum_{y;|y-x|=1} \;
  \psi(y) \;+\;
   V_\omega (x)\psi(x)  ,
\hspace{1cm}
~~\mbox{for} ~~  \psi \in \ell^2(\ZM^d)\,.
\end{equation}

\noindent Here $\lambda$ is a small real parameter providing a measure of the disorder
of the system. The random potential $V_\omega$ is determined
by a family of independent, identically distributed $(iid)$ random variables
$\omega = \left( V_\omega (x)\right)_{x\in\ZM^d}$ with a common
distribution given by a density $g(V)dV$.
We assume that the density $g$ is analytic in a strip about the real axis.
We write $\PM$ for the probability measure generated
by $g$ as an infinite product measure on the probability space $\Omega$.
Precise assumptions and formulation of the results are given in
the next section.

\vspace{.1cm}

In order to describe our results,
let us suppose that we have a family of covariant
observables $A_{\omega,i}$, for $i= 1,
\ldots, N$. A basic example is the $j^{th}$-component of the
velocity operator $V_j = i [ H_\omega, x_j ]$, that is independent of
$\omega$.
The resolvent $R_\omega (z) = (H_\omega - z)^{-1}$ for $H_\omega$ has matrix
elements $G_\omega (x,y; z) =
\langle x | R_\omega (z) | y \rangle$ giving the Green
function. The associated spectral density operator $\rho_\omega (E)$ is given by
$\lim_{\epsilon \rightarrow 0}
\Im (H_\omega -E - i \epsilon )^{-1}$.
The $N$-point correlation function $K_N$, associated with a covariant
family $\{
A_{\omega,j} ~| j=1, \ldots, N \}$, is given by

\begin{equation}
\label{highcorr05.eq-correl1}
K_N (E_1, E_2, \ldots, E_N) \equiv
\int_\Omega d \PM (\omega)
\langle 0 | \rho_\omega (E_1) A_{\omega,1} \rho_\omega (E_2)
A_{\omega, 2} \cdots \rho_\omega (E_N)
A_{\omega, N} | 0 \rangle  .
\end{equation}

\noindent We are interested in the behavior of $K_N$ as a function of the
energies
$E_j, ~j=1, \ldots, N$.
It is not {\it a priori} clear that $K_N$ in (\ref{highcorr05.eq-correl1})
is well-defined.
We will prove, in fact, that $K_N ( E_1, \ldots, E_N)$ is real analytic
in a region of $\RM^N$ away from the
planes where $E_j = E_i$, for $i \neq j$.

As defined in (\ref{highcorr05.eq-correl1}),
the first correlation function $K_1 (E)$, traditionally noted as $\rho
(E)$, is the {\it density
of states} (DOS) of the family $H_\omega$. This function has been 
extensively
studied, especially in one-dimension, and we refer the reader to the
monographs \cite{CL,PasFig} for results on the regularity of the DOS for
one-dimensional lattice models.
In this article, we are especially concerned with lattice models in $d \geq
2$ for which there are very few analyticity results. Constantinescu, 
Fr\"ohlich,
and Spencer \cite{CFS} studied the DOS for lattice models using the random
walk expansion described in section 3. First, these authors proved two 
results
independent of the (nonzero)
disorder. They proved that if $g$ is analytic in a strip of
width $\delta > 4d$, then $\rho (E)$ is analytic in a small strip around
$| \Re E | >> 1$.
They also proved that if $g$ is Gaussian, then $\rho (E)$ is analytic for
$| \Re E | >> 1$, in a region $| \Im E | < (1/ \sqrt{2} ) | \Re E |$.
Furthermore, the DOS decays like a Gaussian as $| E | \rightarrow \infty$
along the real axis.
Note that both of these results are large energy results. Secondly, they
proved that for $g$ Gaussian and large disorder, the DOS $\rho (E)$ is
analytic in a neighborhood of the real axis $\RM$.
Later, Bovier, Campanino, Klein, and Perez \cite{[BCKP]}
obtained stronger smoothness results on the DOS using a supersymmetric
representation of the Green's function. These authors (actually, attributed
to Klein and Perez, unpublished, in \cite{[BCKP]} ) prove two classes of
results of interest here. First, they prove that if
the characteristic function $h$ of the
probability measure satisfies $(1+t)^{(d+k)} h(t) \in L^1 (\RM)$,
then $N(E) \in C^{(k+1)} (\RM)$, for all disorder sufficiently large
(including $k= \infty$). Secondly, if $e^{\alpha t} h(t)$ is bounded for
some $\alpha > 0$, then for any $0 < \alpha_1 < \alpha$, there is a
constant $\lambda_1 > 0$ so that for any $0 \leq \lambda < \lambda_1$,
the IDS $N(E)$ is analytic in a strip $| \Im E| < \alpha_1$.
Our results, restricted to the case $N = 1$, are essentially the same
as this last mentioned result on the analyticity of the DOS in a
strip about the real axis provided the density has a continuation.
Our method of proof is completely different and generalizes
to any higher-order correlation functions.

The higher-order correlation functions have not been studied systematically,
although L.\ Pastur \cite{Pastur}, and one of the authors \cite{Be94,BVESB},
have long emphasized their importance in transport problems.
The second-order correlation functions $K_2 (E_1, E_2)$, for the
choices $A_{\omega, i} = V_i$, are called
the {\it current-current correlation functions}. These functions (actually
measures)
play an important role in the theory of conductivity.
Until now, it was not known if the measure has a density.
Our result on the current-current correlation function is
the first proving the existence of a density for this measure away from the
diagonal. We will comment further on this in section 2. 
The second-order
correlation function has been studied by Pastur and Figotin for a
one-dimensional
quasi-periodic Schr\"odinger operator \cite{FigPas1}. The strong disorder
expansion of the
second moment suggested
by Pastur was developed in Kirsch, Lenoble, Pastur \cite{KLP} providing
insight into
the behavior of the function. In general,
very little is known about this function
and one of our results is that this function is analytic away from
the diagonal $E_1 = E_2$.
The DC conductivity requires control of the two-point correlation function
on the diagonal $E_1 = E_2$. Although we do not achieve this result in this
paper, we give the first results in the study of the correlation functions.


The higher-order correlation functions 
$K_N$ correspond to not-necessarily-positive, bounded, 
Radon measures $K_N (dE_1,
\ldots, dE_N)$. Indeed,
for any $N$, if $f_1, \cdots, f_N$ are 
smooth functions with compact support on
$\RM$ then
\begin{equation}\label{ncorrfnc1}
\mathcal{T}_{\PM}
   \left\{
    \nabla_{j_1} H f_{1}(H) \nabla_{j_2} H f_{2}(H) \cdots \nabla_{j_N} 
   H  f_{N} (H)
   \right\}\;=\;
    \int_{\RM^N }
K_{N} (dE_1, \ldots, dE_{N}) ~f_{1} (E_1) \cdots f_{N}(E_{N})\,  ,
\end{equation}
where $j_k \in \{ 1 , \ldots, d\}$.
The left-hand side of (\ref{ncorrfnc1}) defines tempered distribution in
$N$ variables. Moreover, the left-hand side is
multilinear and bounded with respect to the sup norm of the $f_i$'s
since the operator $\nabla_j H$ is bounded for lattice models.
Therefore $K_N$ belongs to the dual space of $\Cc_0(\RM)^{\otimes N}\simeq
\Cc_0(\RM^N)$, namely it is a bounded Radon measure on $\RM^N$. Except for
the cases $N=1$ and $N=2$, corresponding to the density of  states
and the current-current correlation function, respectively, 
this measure is not necessarily positive. 

Control of higher-order correlation functions
$K_N$ seems to be necessary, for example, in order to
control growth in a dissipative model
describing the Mott variable-range hopping.
The higher-order moments of the position operator $R_j$ are also controlled
by the higher-order correlation functions.
To see this, we note that
\begin{equation}\label{correl1}
R_j(t) - R_j (0) = - i \int_0^t ~ds ~U_s [ H, R_j ] U_s^* =
-i \int_0^t ~ds ~U_s ~\nabla_j H ~U_s^* ,
\end{equation}
where $U_s = e^{-isH}$ and $\nabla_j H = -i [ H , R_j]$ (see
(\ref{derive1})). 
Consequently, there
is a function $F_t (E_1, \ldots, E_{2n})$, depending on $t$, 
so that the $(2n)^{th}$-moment of the position operator $R_j(t) - R_j (0)$,
localized to an energy interval $\Delta$ by projector $P_\Delta$,
may be written as
\begin{eqnarray}\label{correl2}
\mathcal{T}_{\PM} ( P_\Delta | R_j (t) - R_j (0) |^{2n} P_\Delta ) &=&
\int_{\RM^{2(n-1)} \times \Delta }
K_{2n} (dE_1, \ldots, dE_{2n}) ~F_t (E_1, \ldots , E_{2n} ), 
\end{eqnarray}
where
$$ F_t (E_1, \ldots , E_{2n} ) = \prod_{j=1}^{2n} \frac{ 
\sin[(t/2)
(E_j - E_{j-1})] }{ E_j - E_{j-1} }.
$$
Our results on the current-current correlation functions, and the
higher-order correlation functions, are
the first results proving the smoothness of these correlation functions
outside of the coincident planes.


\subsection{Contents of the Paper}

We state the main hypotheses on the models and the main results in section
2. We introduce the random walk expansion of the resolvent in section 3.
In section 4, we give
some basic estimates on Cauchy-type integrals.
To illustrate our method, we apply these to prove the analyticity
of the DOS in section 5. We extend these estimates to multiple
Cauchy-type integrals in section 6. Analyticity of the boundary values of
the Cauchy-type integrals is proved in section 7. The main result on the
analyticity of the correlation
functions is proved in sections 8.

\vspace{.3cm}

\section{The Models, Hypotheses, and the Main Result}
\label{highcorr05.sect-models}

\noindent We now provide precise hypotheses and formulation of our results.
Throughout this paper, we will assume that the single-site probability
density $g$ satisfies the following properties:

\begin{enumerate}
  \item Analyticity: $g$ can be continued as a holomorphic
  function in the strip $\Bb_r = \{ z\in \CM\,; \, |\Im{z}| <r \}$

  \item Boundedness: $\|g\|_r := \sup_{|w|<r} \int_{\RM} dv\, |g(v+\imath
w)| < \infty$.
\end{enumerate}

\noindent
The space of functions fulfilling these conditions will
be denoted by $\Hh_r$ and it will be endowed
with the norm $\|\cdot\|_r$. Note that if $0 < r' \leq r$, then the norm
satisfies $\| g \|_{r'}  \leq \| g \|_r$, for $g \in \Hh_r$.
In addition, it is required that $g$ define a probability distribution
with zero mean and a finite second moment, namely $g(v) \geq 0$,
for $v\in \RM$, and

$$\int_{\RM} dv\, g(v) \;=\; 1\,,
\hspace{1cm}
\int_{\RM} dv\, g(v) v \;=\; 0\,,
\hspace{1cm}
\int_{\RM} dv\, g(v) v^2\; \equiv M_g < \infty . 
$$

\noindent
The measure space in which $\omega$ lives will be denoted by
$\Omega$ and the probability measure defined by the infinite product of
$g(V) dV$
will be denoted $\PM$. Obviously $\ZM^d$ acts on $\Omega$ by
bimeasurable bijective maps and $\PM$ is $\ZM^d$-invariant and ergodic.
A {\em covariant operator} is a family $A=\left( A_{\omega} \right)_{\omega
\in \Omega}$
of bounded operators on $\HG=\ell^2(\ZM^d)$ such that

\begin{enumerate}
\item The map $\omega \in \Omega \mapsto A_{\omega} \in \bs(\HG)$ is
strongly $\PM$-measurable\footnote{Note that the Borel structure of
$\bs(\HG)$ is the same for the weak and
strong topologies since $\HG$ is separable.}.
\item If $T(a)$ denotes the unitary operator of translation by $a$ then
$T(a) A_{\omega} T(a)^{-1} = A_{\trb^a \omega}$, if $\tra^a$ is the action
of $a\in\ZM^d$ on $\Omega$.
\end{enumerate}

\noindent  The Anderson Hamiltonian $H_\omega$, defined in
(\ref{highcorr05.eq-andersonmodel}), is an example of a family of covariant
operators whenever the distribution of the random variables $V_\omega (x)$
has compact support.
However, due to the analyticity condition on $g$, the Hamiltonians
$H_\omega$ are almost surely unbounded and selfadjoint.
It is required then that the field of resolvents
$\omega \in \Omega \mapsto (z\id - H_{\omega})^{-1}\in \bs(\HG)$ be
measurable.
The set of covariant operators is a von Neumann
algebra\footnote{A von Neumann algebra is a \Cs with a predual
\cite{Sakai}.}
~with the pointwise algebraic operations (sum, product, adjoint) and with
the norm
$\| A\| = \essup_{\omega \in \Omega} \|A_{\omega}\|$ \cite{Co94}. An
unbounded, selfadjoint, covariant operator, such as $H_\omega$, is said to
be {\em affiliated to} this algebra if all its spectral projections for
bounded Borel subsets of $\RM$ belong to this algebra. There is a canonical
trace on this algebra given by \cite{Bel02,Co94}

$$\TV (A) \;=\;
   \int_{\Omega} d\PM (\omega) \langle 0|A_{\omega}|0\rangle \;=\;
    \lim_{\Lambda \uparrow \ZM^d}
     \frac{1}{|\Lambda|}
      \TR_{\Lambda} \left( A_{\omega} \right),
\hspace{1cm}
\PM \;\;a.\ e.\
$$

\noindent This von Neumann algebra is denoted by
$L^{\infty}(\TV)$\footnote{For an intrinsic definition without the help of
$\HG$, see \cite{Be93,Co94}}.
~Let $\RV = (R_1,\cdots ,R_d)$ be the {\em position operator} acting on
$\HG$ as a multiplication operator. Then a derivation $\naV$ on the algebra
is defined by

\begin{equation}\label{derive1}
\naV \;=\; (\partial_1, \cdots, \partial_d),
\hspace{1cm} ~\mbox{with}
~~(\partial_{\mu} A)_{\omega} \;=\;
  \imath [R_{\mu}, A_{\omega}],
\end{equation}

\noindent on the dense subalgebra of covariant operators so that the
operators $(\partial_{\mu} A)_{\omega}$ are bounded almost surely. The
Integrated Density of States (IDS) has been proved to satisfy the {\em
Shubin-Pastur formula} (see \cite{Shu79,PasFig,Be86,Be93})

$$\Nn (E) \;=\;
   \lim_{\Lambda \uparrow \ZM^d}
    \frac{1}{|\Lambda |}
     \#\{ \mbox{\small \it eigenvalues of} \;\;
         H_{\omega}\upharpoonright_{\Lambda} \leq E
        \} \;=\;
     \TV (\chi(H\leq E))\qquad a.e.\;\;\omega ,
$$

\noindent
where $\chi(H\leq E)$ denotes the spectral projection of the covariant
Hamiltonian $H_\omega$
on the interval $(-\infty, E]$. The {\em density of states} (DOS) is the
Lebesgue-Stieljes measure $d\Nn$.

\vspace{.1cm}

Transport properties are expressed through the {\em current-current
correlation function}
defined as the positive matrix valued measure on $\RM^2$ given by

\begin{equation}
\label{highcorr05.eq-currentcurrent}
\TV (f(H) \partial_{\nu} H g(H)\partial_{\nu'}H ) \;=\;
   \int_{\RM\times \RM}
    m_{\nu,\nu'}(dE,dE') \; f(E)\, g(E')  ,
\end{equation}

\noindent for $f,g\in \Cc_0(\RM)$. Since the electric current operator
is given by $\JV = (e/\hbar )\naV H$, the measure $m$ is
formally given by $ m_{\nu,\nu'}(dE,dE') = \langle E|J_{\nu}|E' \rangle
\langle E'|J_{\nu'}|E \rangle dE dE'$  (up to a multiplicative constant) in
terms of the matrix elements of the current in the eigenbasis of the
Hamiltonian.

\vspace{.1cm}

One of the main problems for transport in such system is to show
that this measure is absolutely continuous with smooth density.
We prove a
first result in this direction.
Namely, we prove that the density exists and is analytic outside of
the diagonal at all energies in the strong disorder regime.
More precisely, let $\AU = \{A_1, \dots,
A_N\}$ be a finite family of covariant operators as defined above, such as
the velocity operators.
The $N$-point Green function associated with $\AU$ is
defined by

\begin{equation}
\label{highcorr05.eq-Ngreen}
G_{\AU}(\zU) =
   \TV\left(
      (H-z_1)^{-1} A_1 \cdots (H-z_N)^{-1} A_N
      \right)\,,
\hspace{.5cm}
  \zU = (z_1,\cdots,z_N) \in (\CM\setminus \RM)^N\,.
\end{equation}

\noindent Note that the spectral function is a linear combination of two
Green
functions with complex-conjugate imaginary parts.
The DOS is recovered by taking $N=1$, $A_1 = 1$, and taking the imaginary
part. The two-point correlation function is recovered by
taking $N=2$, $A_1=A_2 = \partial_{\nu}H$, expressing the spectral family
in terms of the Green's functions via Stone's formula,
and by considering the
discontinuity of $G_{\AU}(\zU)$ along $z_i = E_i \in \RM$.
The discontinuities are expressed in terms of the boundary values defined as
follows:
let $\sigU = (\sigma_1, \cdots, \sigma_N)\in \{+,-\}^N$, and let $\epsU =
(\epsilon_1,\cdots, \epsilon_N) \in \RM_+^N$.
Then, with the notation $\sigU\cdot \epsU = (\sigma_1\epsilon_1, \cdots,
\sigma_N \epsilon_N)$, we define
the boundary values of (\ref{highcorr05.eq-Ngreen}) as

$$G_{\AU}^{\sigU}(\EU) \;=\;
   \lim_{\epsilon \downarrow 0} G_{\AU}(\EU + \imath \sigU\cdot \epsU)\,.
$$

\noindent The $N$-point correlation functions $K_N$ defined in
(\ref{highcorr05.eq-correl1}) can be expressed as a linear combination of
these boundary values. Our main result is the following theorem.

\begin{theo}
\label{highcorr05.th-smoothG}
Let $H_\omega$ be the Anderson Hamiltonian defined in
(\ref{highcorr05.eq-andersonmodel})
with a distribution $g\in\Hh_r$. Then, given $N \in \NM$, there is $a_0
\equiv a_0 (r,g,d,N) > 0$
such that any family $\AU = \{A_1, \dots, A_N\}$ of covariant operators,
the boundary values of the $N$-point Green function
$G_{\AU}$ defined in (\ref{highcorr05.eq-Ngreen})
are real analytic in domains $\{\EU\in \RM^N\,;\, |E_i-E_j| >
a_0|\lambda|\}$.
\end{theo}

\noindent The proof will use the {\em random walk expansion}
proposed by Fr\"ohlich and Spencer \cite{FrSp}
and used by Constantinescu, Fr\"ohlich, and Spencer \cite{CFS} for the DOS,
which applies in the perturbative domain $|\lambda| \ll 1$ of strong
disorder. As an immediate corollary
we recover the analyticity
result of Bovier, Campanino, Klein, and Perez \cite{[BCKP]},
improving the result of Constantinescu,
Fr\"ohlich, and Spencer \cite{CFS}, for the case
$N=1$. More importantly, we obtain the first results on the existence of a
density for the current-current correlation function.

\begin{coro}
\label{highcorr05.cor-doscurcur}
For the Anderson model, under the assumption of
Theorem~\ref{highcorr05.th-smoothG}, for any $\epsilon > 0$,
there exists a $\lambda_{r,\epsilon} > 0$ so that for $|\lambda| <
\lambda_{r,\epsilon}$,
the DOS is analytic in a strip of width $r-\epsilon > 0$ about the real
axis. Furthermore, there is a constant $0 < a_2 < \infty$ so that
the current-current correlation functions are real
analytic in $(E,E')$ on $\RM^2 \backslash \{ (E, E') ~ | ~ |E - E'| \leq
a_2 |\lambda| \}$.
\end{coro}

\noindent It is not expected that the current-current correlation function
$m (E, E')$ be analytic near the diagonal.
In fact, a perturbative approach \cite{KLP, Pastur}
predicts $m (E_1, E_2) \simeq C| E_1 - E_2 |^2 \ln^{d+1}  | E_1 - E_2 |$,
and a recent result by Klein, Lenoble, and M\"uller \cite{Klein05}
proved an upper bound on the AC conductivity in the spirit of Mott's
formula. This has the form
$\overline{\sigma}_E (\nu) \leq C_0 \nu^2 \log(\frac{1}{\nu})$,
as $\nu \rightarrow 0$, for energies $E$ in the localization
regime at strong disorder. The averaged conductivity $\overline{\sigma}_E
(\nu)$ is defined as $\nu^{-1} \Sigma_E ( [0,\nu])$ for a well-defined
conductivity measure $\Sigma_E$ (see \cite{Klein05}).
If the current-current correlation function has
a density $m(E_1, E_2)$ near the diagonal $E_1 \approx E_2$,
then this result implies
$m (E_1, E_2) \leq C| E_1 - E_2 |^2 \ln^{d+2}  | E_1 - E_2 |$
in the localization regime at strong disorder.
The assumption that such a density exists near the diagonal is
still unproven.
We also have reason to believe that this upper bound might be
supplemented by a similar lower bound, namely

\begin{prob}\label{highcorr05.prob-cp}
Prove that the current-current correlation function is given by a density
that vanishes at $E_1 = E_2$
like $| E_1 - E_2 |^2 \ln^\alpha  | E_1 - E_2 | $,
for some constant $\alpha >0$
depending on the dimension $d$.
\hfill $\Box$
\end{prob}

We mention that this behavior is not consistent with the behavior
predicted by the Mott variable range hopping conductivity argument
\cite{Mott, ES}, which should, in principle,
imply the existence of an essential singularity at coincident energies.
If so, it casts some doubt on the
ability of the one-particle Anderson model to account for the
properties of semiconductors at very low temperatures.
We intend to discuss this behavior for the $2$-point,
and the general $N$-point correlation functions,
in the strong localization regime in a companion article \cite{BH}.

\vspace{.3cm}

\section{Random Walk Expansion of the Resolvent}
\label{highcorr05.sect-randomwalk}

\noindent The method used in these notes is the one proposed in the
early days of the Anderson model by Fr\"ohlich and Spencer \cite{FrSp} and
used to treat the DOS for Gaussian distributions
by Constantinescu, Fr\"ohlich, and Spencer \cite{CFS}.
It is a simple perturbation expansion in the small parameter
$\lambda$ appearing in the definition (\ref{highcorr05.eq-andersonmodel}).
From the definition of $G_{\AU} (\zU)$
in (\ref{highcorr05.eq-Ngreen}), it follows that

\begin{equation}
\label{highcorr05.eq-expansion1}
G_{\AU}(\zU) = \int_{\Omega} d\PM(\omega) \;
     \langle 0 |\frac{1}{H_{\omega}-z_1} |x_1\rangle
      \langle x_1 |A_{1,\omega}|y_1\rangle \; \cdots
        \langle y_{N-1} |\frac{1}{H_{\omega}-z_N} |x_N\rangle
         \langle x_N |A_{N,\omega}|0\rangle ,
\end{equation}

\noindent where repeated coordinates are summed over all $\ZM^d$.
We first show how a general $N$-point function $G_{\AU} (\zU)$ can be
approximated by an $N$-point function constructed from simpler covariant
operators $A$ that we call {\it $r$-monomials}. These are effectively
finite-range operators with analytic coefficients.

\begin{defini}
\label{highcorr05.def-cylind}
Given $r>0$, an element $A\in L^{\infty}(\TV)$ is called an {\em
$r$-monomial}
if there is a family $b_1, \cdots, b_L$ of bounded holomorphic
complex-valued functions in the strip
$\Bb_r = \{z\in \CM\,;\, |\Im{z}|<r\}$, vanishing at infinity,
and a finite set of points $ u_1, \cdots, u_L\in \ZM^d$,
such that the matrix elements of $A_{\omega}$ satisfy
$$
\langle0|A_{\omega}|x\rangle = \prod_{j=1}^L b_j(V(u_j))
\delta_{x,u_0} .
$$
A covariant operator $A$ is an {\em $r$-polynomial}
if it is a finite sum of $r$-monomials.
\end{defini}

\noindent
\noindent
Examples of covariant
operators $A_\omega$ include the Laplacian $A
= \Delta$, the velocity operators $V_j = i[H_\omega, x_j]$,
and inverses of even polynomials in
the random potential $A = V_\omega$ with positive coefficients.
These are not $r$-monomials since the coefficients don't vanish at
infinity. However, they can be
well-approximated by finite linear combinations of $r$-monomials as the
next
proposition shows.
We call a covariant operator $A_\omega$
a {\it finite-range covariant operator} if
there is a finite number $R > 0$
such that if $x,y \in \ZM^d$ with $|x-y| \geq R$, then $\langle
x|A_{\omega}|y\rangle =0$
almost surely.

\begin{proposi}
\label{highcorr05.prop-monomialapprox}
Given a family $\AU$ of $N$ elements in $L^{\infty}(\TV)$,
the $N$-point function $G_{\AU}$ can be approximated uniformly
on any compact subset of $(\CM\setminus \RM)^N$
by a sequence of linear combinations of
$N$-point functions involving only $r$-monomials.
\end{proposi}

\noindent The proof involves the following steps.

\begin{lemma}
\label{highcorr05.lem-covoperapprox}
Let $A$ be an element of the $L^{\infty}(\TV)$. Then,

\noindent
(i) its matrix elements can be written as
$\langle x|A_{\omega}|y\rangle = a(\tra^{-x}\omega, y-x)$ where,
for each $u\in \ZM^d$, the map $a_u: \omega\in \Omega
\mapsto a(\omega,u)$ belongs to $L^{\infty}(\Omega, \PM)$.

\noindent
(ii) the sum $\sum_u |a_u|^2$ converges in $L^{\infty}(\Omega, \PM)$.
\end{lemma}

\noindent  {\bf Proof: } The covariance condition applied to $A$
implies
$\langle x|A_{\omega} |y\rangle = \langle
x-s|A_{\trb^{-s}\omega}|y-s\rangle$
$\PM$-almost surely for all $s\in\ZM^d$. If we define
$a_u ( \omega) = a(\omega,u) \equiv
\langle 0|A_{\omega}|u\rangle$,
then choosing $s=x$ leads to
$\langle x|A_{\omega}|y\rangle = a(\tra^{-x}\omega, y-x)$.
Since $\omega \in \Omega \mapsto A_{\omega}$ is measurable, so are each of
the
maps $a_u$. Since $A$ is bounded, it follows that
$\|a_u\|_{L^{\infty}} \leq \|A\|$.
Hence $a_u \in L^{\infty}(\Omega, \PM)$. In the same way,
one shows that
$\langle 0 |A_{\omega} A_{\omega}^{\ast} |0\rangle =
\sum_{u\in\ZM^d} |a_u(\omega)|^2 \leq \|A\|^2$,
showing that this sum actually converges in $L^{\infty}(\Omega, \PM)$.
\hfill $\Box$

\begin{lemma}
\label{highcorr05.lem-rpoly}
Any element $A\in L^{\infty}(\TV)$ can be weakly
approximated by a sequence of $r$-polynomials.
\end{lemma}

\noindent
{\bf Proof: } Thanks to Lemma~\ref{highcorr05.lem-covoperapprox}(ii),
$A$ can be uniformly approximated by a finite range operator,
namely given $\epsilon >0$, there is $N\in \NM$ such that
$\sum_{|u|>N} |a_u|^2 \leq \epsilon^2$.
Hence setting $a^{(N)}_u = a_u$, if $|u|\leq N$, and  $a^{(N)}_u = 0$,
otherwise, we obtain a finite range operator
$A_N \in L^{\infty}(\TV)$ such that $\|A-A_N\|\leq \epsilon$.
On the other hand, since $\Omega$ can be taken as the Cartesian
product $\RM^{\ZM^d}$, any element of $L^{\infty}(\Omega, \PM)$
can be weakly approximated (for the weak-$\ast$ topology) by a
continuous cylindrical function. A function $f:\Omega\mapsto \CM$
is called cylindrical if there is $L\in \NM_{\ast}$,
a continuous function $F\in \Cc_0(\RM^L)$, vanishing at
infinity, and  a finite subset $\{x_1, \cdots, x_L\}\subset \ZM^d$
such that $f(\omega) = F(V_\omega(x_1), \cdots, V_\omega(x_L))$.
Since $\Cc_0(\RM^L)$ is the uniform closure of the algebraic tensor
product $\Cc_0(\RM)^{\otimes L}$ such a function $F$ can be uniformly
approximated by a finite sum of functions of the form
$f_1(V_\omega(x_1))\cdots f_L(V_\omega(x_L))$,
where $f_i\in \Cc_0(\RM)$. At last,
the space $\Hh_r$, made of functions in $\Cc_0(\RM)$ that can be
continued as holomorphic functions on the strip $\Bb_r$ vanishing at
infinity,
is dense in $\Cc_0(\RM)$.
\hfill $\Box$

\vspace{.1cm}

\noindent We let $\Hh$ be the space of functions on $(\CM\setminus\RM)^N$
that
are holomorphic and bounded at infinity, and
endowed with the topology of uniform convergence on compact sets.

\begin{lemma}
\label{highcorr05.lem-Nptweakcont}
The map $\AU = (A_1, \cdots , A_N) \in L^{\infty}(\TV)^{\times N}
\mapsto G_{\AU}\in \Hh$ is multilinear and weak-$\ast$ continuous.
\end{lemma}

\noindent
{\bf Proof: }
This is a consequence of the GNS-representation theorem and the
property that on a von Neumann algebra
the weak and the strong topology coincide on bounded sets.
\hfill $\Box$

\vspace{.2cm}
\noindent
{\bf Proof of Proposition~\ref{highcorr05.prop-monomialapprox}: }
It is a consequence of the previous
Lemmas~\ref{highcorr05.lem-covoperapprox}, \ref{highcorr05.lem-rpoly}
and \ref{highcorr05.lem-Nptweakcont}.
\hfill $\Box$

\vspace{.2cm}

\noindent We now return to the expansion (\ref{highcorr05.eq-expansion1}).
Since the kinetic term in (\ref{highcorr05.eq-andersonmodel}) is given by
the discrete Laplacian (without diagonal term), the Green function can then
be expanded in formal power series in $\lambda$. This gives

$$\langle y |\frac{1}{H_{\omega}-z} |x\rangle = \sum_{j=0}^\infty
(-\lambda)^j \langle y| \frac{1}{V - z} \left[ H_0 \frac{1}{V-z} \right]^j
| x \rangle\,.
$$

\noindent We next use the fact that the Laplacian $H_0$ couples only
nearest-neighbor terms to obtain a path expansion of matrix elements of the
resolvent. We need some notation. We denote by $\gamma$ a {\em path} from
$y$ to $x$, namely it
is a sequence $(x_0=y, x_1, \cdots, x_{n-1}, x_n=x)$ where $x_k\in \ZM^d$
for all $k$'s and $|x_k-x_{k-1}| =1$, for $1\leq k \leq n$. Note that all
the points need not be distinct. The points $x_0$ and $x_n$ are called the
{\em initial} and the {\em final} points of $\gamma$ and will be denoted by
$\partial_0 \gamma$ and $\partial_1\gamma$, respectively. We call $n
=|\gamma|$ the {\em length} of $\gamma$. We denote by $\Vv(\gamma)=
\{x_k\,;\, 0\leq k\leq n \}$
the family of (distinct) {\em vertices} of $\gamma$\footnote{If the path
$\gamma$ passes through the same vertex more than once, $\Vv(\gamma)$ has
less than $n+1$ elements in general.}. It will be convenient to denote by
$\#\gamma$ the cardinality of $\Vv(\gamma)$so that $\#\gamma \leq n+1$. The
path expansion takes the form:

\begin{equation}
\label{highcorr05.eq-rwegreen}
\langle y |\frac{1}{H_{\omega}-z} |x\rangle \;=\;
   \sum_{\gamma :y\mapsto x}
    (-\lambda)^{|\gamma|}\;
     \prod_{k=0}^n
      \frac{1}{V(x_k)-z}\,.
\end{equation}

\noindent Given $u\in \ZM^d$, let $n_{\gamma}(u)$ be the number of $k\in
[0,n]$
such that $x_k=u$, so that $n+1 = |\gamma| + 1 = \sum_{u\in \ZM^d}
n_{\gamma}(u)$.
Hence (\ref{highcorr05.eq-rwegreen}) can be written as

\begin{equation}
\label{highcorr05.eq-rwegreen2}
\langle y |\frac{1}{H_{\omega}-z} |x\rangle \;=\;
   \sum_{\gamma :y\mapsto x}
    (-\lambda)^{|\gamma|}\;
     \prod_{u\in \ZM^d}
      \frac{1}{(V(u)-z)^{n_{\gamma}(u)}}\,.
\end{equation}

\noindent This formula will be used to expand $G_{\AU}(\zU)$ given in
(\ref{highcorr05.eq-expansion1}). Since $G_{\AU}(\zU)$ contains $N$ such
Green functions, this expansion
will require $N$ paths, namely a family $\Gamma$ of $N$ paths,
$\Gamma = (\gamma_1, \cdots, \gamma_N)$. For such a family, the following
notation will be used: $|\Gamma| = \sum_{i=1}^N |\gamma_i|$,
$\Vv(\Gamma) = \cup_{i=1}^N \Vv(\gamma_i)$, $\# \Gamma$ is the
cardinality of $\Vv(\Gamma)$, while if $u\in \ZM^d$, $n_{\Gamma}(u) =
\sum_{i=1}^N n_{\gamma_i}(u)$. The initial and final points of $\Gamma$
are defined by $\partial_0\Gamma = \partial_0 \gamma_1$ and
$\partial_1 \Gamma = \partial_1 \gamma_N$, respectively.
An $N$-path is $\AU$ compatible if $\langle \partial_1\gamma_i |A_{i,\omega}
|\partial_0\gamma_{i+1} \rangle \neq 0$, for $i\in [1,N]$,
with the convention $\gamma_{N+1} = \gamma_1$.
Let then $\ps(\AU)$ be the set of $\AU$-compatible $N$-paths
with $\partial_0 \Gamma= \partial_1\Gamma =0$.
Thanks to (\ref{highcorr05.eq-rwegreen2}),
the $N$-point correlation can be written as

$$
G_{\AU}(\zU) =
   \int_{\Omega} d\PM(\omega) \;
    \sum_{\Gamma\in \ps(\AU)}
     (-\lambda)^{|\Gamma|}
      \prod_{u\in \ZM^d}
       \prod_{i=1}^N
        \frac{1}{(V(u)-z_i)^{n_{\gamma_i}(u)}} \;
         \prod_{i=1}^N
  \langle \partial_1 \gamma_i |A_{i,\omega}|\partial_0 \gamma_{i+1}
\rangle\,.
$$

\noindent
Let the operators $A_{i,\omega}$ be all $r$-monomials.
It follows from Definition \ref{highcorr05.def-cylind} and Lemma
\ref{highcorr05.lem-covoperapprox}
that their matrix elements factorize according to

$$\langle x |A_{i,\omega}|y\rangle \;=\;
   \delta_{y-x,u_i} \,
    \prod_{u\in\ZM^d} a_{i,u-x}(V(u))\,,
$$

\noindent for some $u_i \in \ZM^d$
and where $a_{i,u} = 1$ for all but a finite number of indices $(i,u)$.
It follows that an $N$-path is $\AU$-compatible if and only if

\begin{equation}
\label{highcorr05.eq-compt1}
\Gamma \in \ps(\AU)\;\;
   \Longleftrightarrow\;\;
    \partial_0\gamma_{i+1}-\partial_1 \gamma_i = u_i \;\;
    \forall i\in [1,N]\,.
\end{equation}

\noindent Hence there is no need to insert the product of the
Kronecker symbols associated with the matrix elements of the
$A_{i,\omega}$'s.
Then, since the probability measure $\PM$ factorizes with respect to the
sites, the previous formula becomes:

$$G_{\AU}(\zU) =
   \sum_{\Gamma\in \ps(\AU)}
    (-\lambda)^{|\Gamma|}
   \prod_{u\in \ZM^d}
    \int_{-\infty}^{+\infty} dv\; g(v)
     \prod_{i=1}^N
      \frac{
       a_{i,u-\partial_1 \gamma_i}(v)
            }{(v-z_i)^{n_{\gamma_i}(u)}}\,.
$$

\noindent As in (\ref{highcorr05.eq-Jn}) of
Section~\ref{highcorr05.sect-Ncauchy},
if $\nU = (n_1, \cdots, n_N)\in \NM^N$ and $\zU = (z_1, \cdots, z_N)\in
(\CM\setminus \RM)^N$,
it is convenient to set

\begin{equation}
\label{highcorr05.eq-cauchyintegral1}
J_{\nU}(h;\zU) \;=\;
   \int_{-\infty}^{+\infty} dv\; h(v)
     \prod_{i=1}^N
      \frac{1}{(v-z_i)^{n_i+1}}\,,
\end{equation}

\noindent Using the notation $\nU_{\Gamma}(u) = (n_{\gamma_1}(u),
\cdots, n_{\gamma_N}(u))$ and $\fU = (1,1,\cdots,1)$,
the previous formal $N$-path expansion becomes

\begin{equation}
\label{highcorr05.eq-Nptrwe}
G_{\AU}(\zU) =
\sum_{\Gamma\in \ps(\AU)}
  (-\lambda)^{|\Gamma|}
     \prod_{u\in \ZM^d}
      J_{\nU_{\Gamma}(u)-\fU}(g_{\Gamma,u};\zU)\,,
\end{equation}

\noindent where

$$g_{\Gamma,u}(v) \;=\;
   g(v) \prod_{i=1}^N
    a_{i,u-\partial_1 \gamma_i}(v)\,.
$$

\noindent
As a final result in this section, we estimate the number of
$\AU$-compatible $N$-paths of a given length.

\begin{lemma}
\label{highcorr05.lem-countingpath}
Let $\AU$ be a family of $N$ r-monomials.
Then the number of $\AU$-compatible $N$-paths of total length $n$
is bounded from above by

$$
\# \{\Gamma \in \ps(\AU)\,;\, |\Gamma|=n\} \;\leq \;
    (2d)^n\,.
$$
\end{lemma}

\noindent  {\bf Proof: } The initial point of $\Gamma$ is fixed at $x=0$.
There are $2d$ ways of choosing a neighbor of $0$.
This gives the first vertex $x_1$ of $\gamma_1$.
If $x_1, x_2, \cdots, x_j$ have been chosen, there are again $2d$
neighbors of $x_j$ giving $2d$ admissible choices for $x_{j+1}$.
Hence there are exactly $(2d)^{|\gamma_1|}$ ways of choosing $\gamma_1$.
Assume $\gamma_1, \cdots, \gamma_k$ have been chosen.
Then, the final point $\partial_1\gamma_k$ is fixed so that the
$\AU$-compatibility fixes the initial point $\partial_0\gamma_{k+1}$
unambiguously due to (\ref{highcorr05.eq-compt1}).
The same argument shows that the number of possible choices
for $\gamma_{k+1}$ is at most $(2d)^{|\gamma_{k+1}|}$.
This leads to the result by recursion on $k\in [1,N]$.
\hfill $\Box$

\vspace{.3cm}

\section{Estimates on Cauchy-type Integrals}
\label{highcorr05.sect-cauchy}

\noindent In order to prove analyticity of the correlation functions
$G_{\AU} (
\zU)$, as expressed in (\ref{highcorr05.eq-Nptrwe}),
we first need to analyze the Cauchy-type integrals $J_{\nU} (h; \zU)$
defined in (\ref{highcorr05.eq-cauchyintegral1}).
We begin with estimates on the simplest form of these integrals for
which $\underline{n}$ and $\underline{z}$ depend on one variable only.
We will treat the general case in section 6 after we apply the results of
this section to the DOS.
The first result on the behavior
of the density $g$ is the following

\begin{lemma}
\label{highcorr05.lem-derivative}
Let $g\in\Hh_r$ and let $g^{(n)}$ be its $n^{th}$ derivative, then for any
$0 < \rho < r$, we have

$$\sup_{z:|\Im z|\leq r-\rho}|g(z)| \;\leq\;
   \frac{1}{\pi \rho}\|g\|_r\,,
\hspace{2cm}
\|g^{(n)}\|_{r-\rho}\; \leq \;
   \frac{n!}{\rho^n}
    \|g\|_r\,.
$$
\end{lemma}

\noindent
{\bf Proof:} Let $z \in\Bb_{r-\rho} = \{ z \in \CM ~| ~| \Im z| < r - \rho
\}$,
and let $\gamma$ denote a path contained in the strip $\Bb_r$ homotopic to
the circle centered at $z$ of radius $\rho$. Thanks to the Cauchy formula

$$g(z) \;=\;
   \oint_{\gamma}\frac{dz'}{2\imath \pi} \frac{g(z')}{(z'-z)}\,.
$$

\noindent Taking $\gamma$ as the union of the lines $\gamma_\pm = \{ z+u\pm
\imath \rho\,;\, u\in \RM\}$ this gives

$$|g(z)| \;=\; \left|
   \int_{-\infty}^{+\infty} \frac{du}{2\imath \pi}
     \left\{
      \frac{g(u+z-\imath \rho)}{u-\imath \rho} -
       \frac{g(u+z+\imath \rho)}{u+\imath \rho}
     \right\} \right| \;\leq\;
   \frac{1}{\pi\rho} \|g\|_r\,.
$$

\noindent Using again the Cauchy formula, with now $\gamma$ being the circle
centered at $z$ of radius $\rho$, gives

$$g^{(n)}(z) \;=\; n!
   \oint_{\gamma} \frac{dz'}{2\imath \pi} \frac{g(z')}{(z'-z)^{n+1}}\;=\;
    \frac{n!}{\rho^n}
     \int_0^{2\pi} \frac{d\theta}{2\pi}
        g(z+\rho e^{\imath \theta}) e^{-\imath n\theta}\,.
$$

\noindent Integrating the absolute values of both sides over a line parallel
to the real axis gives the result.
\hfill $\Box$

\vspace{.1in}

\noindent
Let $I_n(g;z)$ be defined by

\begin{equation}
\label{highcorr05.eq-cauchyintegral2}
I_n(g;z) \;=\;
   \int_{-\infty}^{+\infty} dv \;
    \frac{g(v)}{(v-z)^{n+1}}\,,
\end{equation}

\noindent which is convergent for $\Im z \neq 0$. We have the following 
identities.

\begin{lemma}
\label{highcorr05.lem-In}
If $g\in\Hh_r$, then for any $z \in \CM$ with $\Im z \neq 0$,

$$I_n(g;z) \;=\;
   \frac{1}{n!} \frac{d^n}{dz^n} I_0(g;z) \;=\;
    \frac{1}{n!} I_0(g^{(n)};z)\,.
$$
\end{lemma}

\noindent
{\bf Proof: } The first identity is a direct consequence of the definition.
For the second equality, we note that

$$\frac{dv}{(v-z)^{n+1}} = -\frac{1}{n} \, d\frac{1}{(v-z)^n}\,.
$$

\noindent Since $g\in\Hh_r$, it follows from an integration by parts that

$$I_n(g;z) =
   -\frac{1}{n} \frac{g(v)}{(v-z)^n}\upharpoonright_{-\infty}^{+\infty}
\;+\;
     \frac{1}{n}
      \int_{-\infty}^{+\infty} dv \;
       \frac{g^{(1)}(v)}{(v-z)^n}\,.
$$

\noindent The first term vanishes while the second gives $I_n(g;z)=
I_{n-1}(g^{(1)};z)/n$. The formula follows by recursion.
\hfill $\Box$

\begin{lemma}
\label{highcorr05.lem-Izero}
If $g\in \Hh_r$, then

\begin{equation}
\label{highcorr05.eq-bv1}
\lim_{\epsilon \downarrow 0}
   I_0(g,E \pm \imath \epsilon) \;=\;
    \int_0^{\infty} du\; \frac{g(E+u)-g(E-u)}{u} \;\pm\;
     \imath \pi g(E)\,.
\end{equation}

\end{lemma}

\noindent
{\bf Proof: } Thanks to Cauchy's formula,
this limit can be computed by using a deformed path $\gamma$ avoiding
the point $z=E$. A possible choice, for positive imaginary part,
is $\gamma = \gamma_- \cup \gamma_0 \cup \gamma_+$,
where $\gamma_- =(-\infty , \epsilon]$, $\gamma_0 =
\{\epsilon e^{\imath \theta}\,;\, -\pi \leq \theta \leq 0\}$,
and $\gamma_+= [\epsilon, +\infty)$. This gives a decomposition of
$I_0(g;E+\imath 0)$ into three integrals $I_-+I_0+I_+$.
After the change of variable $v= E\pm u$,
the contributions of $\gamma_\pm$ are given by

$$I_- \;=\;
   -\int_{\epsilon}^{\infty} \frac{du}{u}\; g(E-u)\,,
\hspace{1cm}
   I_+ \;=\;\int_{\epsilon}^{\infty} \frac{du}{u}\; g(E+u)\,.
$$

\noindent Setting $v=E+\epsilon e^{\imath \theta}$ gives

$$I_0 \;=\;
   \int_{-\pi}^0 \imath d\theta \; g(E+\epsilon e^{\imath \theta})
    \;\stackrel{\epsilon \downarrow 0}{\longrightarrow}\;
     \imath \pi g(E)\,.
$$

\noindent
Since $g(E+u)-g(E-u)$ vanishes like $\mathcal{O} (u)$ for $u\rightarrow 0$,
the sum $I_+ + I_-$ converge as $\epsilon \rightarrow 0$ giving the result.
\hfill $\Box$

\begin{lemma}
\label{highcorr05.lem-Izeroestim}
For $r > 0$, if $g\in \Hh_r$, then

$$\sup_{E\in \RM} |I_0(g;E\pm \imath 0)| \; \leq \;
   \left(
    (\frac{8}{\pi}+ 2)\frac{1}{r^2}+ \frac{1}{r} +1
   \right) \|g\|_r .
$$
\end{lemma}

\noindent
{\bf Proof: } Using the formula given in Lemma~\ref{highcorr05.lem-Izero},
the integral (\ref{highcorr05.eq-bv1}) over $u$  decomposes into $\int_0^1 
du (\cdot) +
\int_1^{\infty} du(\cdot)$.
For the first integral, we write $g(E+u) -g(E-u) = \int_{-u}^u g'(E+x) ~dx$,
and
integrate by parts in the variable $u$, to obtain,
\begin{eqnarray*}
\int_0^{1} du\; \frac{g(E+u)-g(E-u)}{u} &=&
\int_0^1 \frac{du}{u} \left[ \int_{-u}^{+u} dx\; g^{(1)} (E+x) \right] \\
&=& \int_0^1 dx \ln{(1/x)} \left(g^{(1)}(E+x)+ g^{(1)}(E-x)\right)\,.
\end{eqnarray*}

\noindent Another integration by parts, gives

\begin{eqnarray}
\label{highcorr05.eq-pp}
\int_0^{1} du\; \frac{g(E+u)-g(E-u)}{u}
&=&
   \left(
    g^{(1)}(E+1)+ g^{(1)}(E-1)
   \right)\nonumber\\
&& -\int_0^1 dx \left(g^{(2)}(E+x)- g^{(2)}(E-x)\right)(x-x\ln{x}).
\end{eqnarray}

\noindent To bound this, we use Lemma~\ref{highcorr05.lem-derivative}, and
obtain,

$$|g^{(1)}(E')| \leq \frac{1}{\pi (r-\rho)} \|g^{(1)}\|_{r-\rho}
   \leq
    \frac{1}{\pi \rho (r-\rho)}\|g\|_r\,,
\hspace{2cm}
\forall E'\in\RM,\;\; 0<\rho<r\,.
$$

\noindent The choice $\rho =r/2$ gives the optimal bound on the right side,
so that

$$
\sup_{E'\in \RM}  | g^{(1)}(E')|\leq
   \frac{4}{\pi r^2} \|g\|_r\,.
$$

\noindent Since $0\leq (x-x\ln{x}) \leq 1$, for $0\leq x\leq1$,
the second term in (\ref{highcorr05.eq-pp})
is dominated by $\|g^{(2)}\|_{r'}$ for all $0\leq r'\leq r$, so that, using
Lemma~\ref{highcorr05.lem-derivative} again,

$$\left|
   \int_0^{1} du\; \frac{g(E+u)-g(E-u)}{u}
  \right| \leq
   (\frac{8}{\pi}+ 2)\frac{1}{r^2} \|g\|_r\,.
$$

\noindent The second part is simply dominated by

$$\left|
   \int_1^{\infty} du\; \frac{g(E+u)-g(E-u)}{u}
  \right| \leq \|g\|_r\,.
$$

\noindent Thanks to Lemma~\ref{highcorr05.lem-derivative} again, the last
term on the right in (\ref{highcorr05.eq-bv1})
is bounded by

$$\left| \imath \pi g(E)\right| \leq
   \frac{1}{r} \|g\|_r\,.
$$

\noindent Overall, this gives

$$|I_0(g;E\pm \imath 0)| \;\leq\;
   \left(
    (\frac{8}{\pi}+ 2)\frac{1}{r^2}+ \frac{1}{r} +1
   \right) \|g\|_r .
$$
\hfill $\Box$

\vspace{.3cm}

\section{Analyticity of the Density of States}
\label{highcorr05.sect-dos}

\noindent In order to illustrate our technique in the simplest setting, we
prove the analyticity of the density of states (DOS) in the strong
disorder regime. This is a new and different
proof of a result in \cite{[BCKP]}
obtained that was obtained
using the supersymmetric replica
trick for the Green's function.
It is an improvement of the result of Constantinescu, Fr\"ohlich, and
Spencer \cite{CFS}.
The case $N=1$ requires only the Cauchy-type
integral estimates of section \ref{highcorr05.sect-cauchy}.
The DOS exists as a function in $L^1_{loc} ( \RM )$ for lattice models since
the integrated
density of states is globally Lipschitz continuous.
The main result of this section is the following theorem.

\begin{theo}
\label{highcorr05.th-dos1}
Let $H_\omega$ be the Anderson Hamiltonian defined in
(\ref{highcorr05.eq-andersonmodel})
with a distribution $g\in\Hh_r$, for some $r > 0$.
Then, for any $0 < \epsilon < r$, there is $\lambda_{r, \epsilon} > 0$
such that for all $0 \leq \lambda < \lambda_{r, \epsilon}$,
the DOS for $H_\omega$ is analytic in a strip of width $r-\epsilon > 0$
about the real axis.
\end{theo}

\noindent
In order to prove Theorem \ref{highcorr05.th-dos1}, we recall that the DOS
$\rho (E)$ is
given by

\begin{eqnarray}
\label{highcorr05.eq-dos11}
\rho (E) &=&
\int_\Omega d \PM ~~\langle 0 | \rho_\omega (E) |0 \rangle
  \nonumber \\
&=& \EM \{ \langle 0 | \Im G(E + i 0 ) | 0 \rangle \} \nonumber \\
&=& \lim_{\epsilon \rightarrow 0} \frac{1}{2i}
     \left\{
      G_{\underline{1} } (E + i \epsilon ) -
      G_{\underline{1} } (E - i \epsilon )
     \right\}\,,
\end{eqnarray}

\noindent using the notation of (\ref{highcorr05.eq-Ngreen}), and
$\rho_\omega (E)$ is the spectral function for $H_\omega$ introduced in
(\ref{highcorr05.eq-correl1}). We consider functions the $G_{\underline{1}}
(E \pm i \epsilon) \equiv G_{\underline{1}}^\pm (E)$ appearing in
(\ref{highcorr05.eq-dos11}). We prove that each function is real analytic on
$\RM$ and that each has a holomorphic continuation to the lower,
respectively, upper, half complex plane in a strip of size $r- \epsilon$,
provided $| \lambda |$ is small enough. Using the random walk expansion
described in section 2, we find

\begin{equation}
\label{highcorr05.eq-dos2}
G_{\underline{1}}(z) = \sum_{\gamma: 0 \rightarrow 0} (- \lambda)^{|
\gamma| } \prod_{u \in \ZM^d} I_{n_\gamma (u)-1} (g; z ) ,
\end{equation}

\noindent where the Cauchy-type integral $I_{n_\gamma (u)} (g; z )$ is given
by

$$I_{n} (g;z) \equiv
   \int_{-\infty}^\infty dv  \;
    \frac{g(v)}{( v - z)^{n+1}}\,,
$$

\noindent as defined in (\ref{highcorr05.eq-cauchyintegral2}). With
reference to Section~\ref{highcorr05.sect-Ncauchy}, we note that

\begin{equation}
\label{highcorr05.eq-doscauchy1}
J_n^{\pm} (g;E)
\equiv
  \lim_{\epsilon \rightarrow 0}
   \frac{1}{n!} I_0 ( g^{(n)}; E \pm i \epsilon )\,.
\end{equation}

\begin{lemma}
\label{highcorr05.lem-lemmados1}
The function $J_n^{\pm} (g;E)$, defined in (\ref{highcorr05.eq-doscauchy1}),
is analytic on $|\Im E| < r$ and its derivatives have the form

\begin{equation}
\label{highcorr05.eq-derivdos1}
\frac{1}{l!} \frac{\partial^l}{\partial E^l} J_n^{\pm} (g; E) =
\frac{ (n+l)!}{n! l!} I_0^{\pm} \left( \frac{g^{(n+l)}}{(n+l)!} ; E \right),
\end{equation}

\noindent where

$$
I_0^{\pm} (h;E) \equiv \lim_{\epsilon \rightarrow 0} I_0 (h; E \pm i
\epsilon ) .
$$
\end{lemma}

\noindent {\bf Proof.} The formula (\ref{highcorr05.eq-derivdos1}) follows
from Lemma~\ref{highcorr05.lem-In} and Lemma~\ref{highcorr05.lem-Izero}. The
analyticity follows from the hypotheses on $g$ and the explicit formula in
Lemma~\ref{highcorr05.lem-Izero}.
\hfill $\Box$

\vspace{.2cm}

\noindent In order to prove real analyticity, we prove that the sum of terms
on the left side of  (\ref{highcorr05.eq-derivdos1}) converges uniformly.
For this it is sufficient to estimate the sum

\begin{equation}
\label{highcorr05.eq-normdefn1}
\| J_n^{\pm} (g; \cdot) \|_\delta \equiv \sum_{l=0}^\infty
\frac{\delta^l}{l!}
\sup_{E \in \RM}
  \left|
    \frac{\partial^l }{\partial E^l} J_n^{\pm} (g; E)
  \right|\,,
\end{equation}

\noindent
for $0 < \delta < r$. Since the diameter $\delta$ will be shown to be
independent of the energy, it follows that the sum converges to an analytic
function for $|\Im E| < \delta$.

\begin{lemma}
\label{highcorr05.lem-lemmados2}
For $0 < \delta < r$, the following estimate holds

$$ \| J_n^{\pm} (g; \cdot) \|_\delta \leq
    \frac{ r e^2 C}{4} \frac{(n+3)^2}{(r - \delta)^{n+3}}\; \|g\|_r\,,
$$

\noindent with $C = (8/\pi) + 2+ r +r^2$.
\end{lemma}

\noindent {\bf Proof.} Lemma~\ref{highcorr05.lem-Izero}, Lemma~\ref
{highcorr05.lem-Izeroestim} and then Lemma~\ref{highcorr05.lem-derivative}
lead to

$$\| I_0^{\pm} \left( \frac{ g^{(l+n)} }{ (n+l)!}\,;\, \cdot  \right)
\|_\infty
   \leq \frac{1}{r_1^{n+l}}
    ~\frac{C \|g \|_r}{(r-r_1)^2}\,,
$$

\noindent for any $0 < r_1 < r$. Summing over $l$ and using
formula~(\ref{highcorr05.eq-taylor})
leads to a convergent sum provided $\delta < r_1<r$ such that

\begin{equation}
\label{highcorr05.eq-supnorm2}
\| J_n^{\pm} (g; \cdot) \|_\delta \leq \frac{r_1}{(r_1 - \delta)^{n+1}}
~\frac{C \|g \|_r}{(r-r_1)^2} .
\end{equation}

\noindent Maximizing the denominator on the right side of
(\ref{highcorr05.eq-supnorm2}) over $r_1$, gives the result.
\hfill $\Box$

\vspace{.2cm}

\noindent {\bf Proof of Theorem 2.} Anticipating
Section~\ref{highcorr05.sect-realanal}, Lemma~\ref{highcorr05.lem-Eealg}
implies that the norm $\|\cdot\|_\delta$, defined in
eq.~(\ref{highcorr05.eq-normdefn1}), is multiplicative. Thus, from the path
expansion (\ref{highcorr05.eq-dos2}), it follows that

$$\| G_{\underline{1}}^\pm ( \cdot ) \|_\delta \leq
   \sum_{\gamma:0 \rightarrow 0}
    |\lambda|^{|\gamma|}
     \prod_{u\in \ZM^d}
       \| J_{n_\gamma(u)-1}^\pm\|_\delta\,.
$$

\noindent The bound $(n/2+1)^2 \leq e^n$ implies $ \prod_{u \in \mathcal{V}
(\gamma)} ( n_\gamma (u)/2  + 1 ) \leq e^{| \gamma|}$. Definition
(\ref{highcorr05.eq-normdefn1}) and Lemma \ref{highcorr05.lem-lemmados2}
lead to

$$\| G_{\underline{1}}^\pm ( \cdot ) \|_\delta \leq
   \sum_{\gamma:0 \rightarrow 0} | \lambda|^{|\gamma|}
    \left[
     \frac{re^2C\|g\|_r}{(r-\delta)^2}
    \right]^{|\Vv(\gamma)|}
     \; \frac{e^{|\gamma|}}{(r-\delta)^{|\gamma|}}\,.
$$

\noindent Using the inequality $| \mathcal{V} (\gamma) | \leq | \gamma| + 1$
this leads to

$$\| G_{\underline{1}}^\pm ( \cdot ) \|_\delta \leq
  \frac{re^2C\|g\|_r}{(r-\delta)^2}
   \sum_{\gamma:0 \rightarrow 0}
    \left[
     \frac{|\lambda|r e^3 C\|g\|_r}{(r-\delta)^3}
    \right]^{|\gamma|}\,.
$$

\noindent Finally, from Lemma \ref{highcorr05.lem-countingpath}, the number
of paths of length $n$ is bounded by $(2d)^n$ so the sum over paths can be
changed into a sum over $n$ to get

$$\| G_{\underline{1}}^\pm ( \cdot ) \|_\delta \leq
  \frac{re^2C\|g\|_r}{(r-\delta)^2}
   \sum_{n\geq 0}
    \left[
     \frac{2d|\lambda|re^3 C\|g\|_r}{(r-\delta)^3}
    \right]^{n}\,.
$$

\noindent So for any small $\epsilon > 0$, if $\lambda_{r,\epsilon} \equiv
\epsilon^3 / ( 2d  e^3 rC\|g\|_r  )$, then for any $|\lambda| <
\lambda_{r,\epsilon}$, the DOS is analytic in the strip $| \Im E | < r -
\epsilon$.
\hfill $\Box$

\vspace{.3cm}

\section{Estimates on $N$-point Cauchy-type Integrals}
\label{highcorr05.sect-Ncauchy}

\noindent In this section, the general $N$-point Cauchy-type integrals
appearing in
(\ref{highcorr05.eq-cauchyintegral1})--(\ref{highcorr05.eq-Nptrwe}) is
considered. Results similar to those of section 3 for the case of $N=1$ will
be obtained. Let $\nU = (n_1,\cdots, n_N)\in \NM^N$, and let $\zU =
(z_1,\cdots, z_N)\in (\CM\setminus\RM)^N$. The following integral will be
considered

\begin{equation}
\label{highcorr05.eq-Jn}
J_{\nU}(g;\zU) \;=\;
\int_{-\infty}^{+\infty} dv\;
  \frac{g(v)}{(v-z_1)^{n_1+1}\cdots (v-z_N)^{n_N+1}}\,,
\end{equation}

\noindent with the following usual convention for multi-indices

$$\nU ! = \prod_{k=1}^N n_k !\,,
\hspace{1cm}
   |\nU | = \sum_{k=1}^N n_k\,,
\hspace{1cm}
    \partial_{\zU}^{\nU} =\prod_{k=1}^N
     \frac{\partial^{n_k}}{\partial z_k^{n_k}} .
$$

\begin{lemma}
\label{highcorr05.lem-Jn1}
The following formul{\ae} hold
\begin{equation}
\label{highcorr05.eq-Jn1}
J_{\nU}(g;\zU) \;=\;
   \frac{1}{\nU !} \; \partial^{\nU} \; J_{\OU}(g;\zU) ,
\end{equation}

\noindent and

\begin{equation}
\label{highcorr05.eq-Jn11}
J_{\OU}(g;\zU) \;=\;
\sum_{i=1}^N
  \prod_{j:\neq i}
   \frac{1}{z_i -z_j} \;
    I_0(g;z_i)
\end{equation}
\end{lemma}

\noindent
{\bf Proof:} The first formula~(\ref{highcorr05.eq-Jn1}) is obtained by
using repetitively the identity

\begin{equation}
\label{highcorr05.eq-der1survn}
\frac{1}{(v-z)^{n+1}} \;=\;
   \frac{1}{n!}\; \frac{d^n}{dz^n}\;
    \frac{1}{v-z}\,.
\end{equation}

\noindent
The formula~(\ref{highcorr05.eq-Jn11}) is obtained from the following
identity,
valid if $z_k\neq z_l$ for $k\neq l$

\begin{equation}
\label{highcorr05.eq-primefactor}
\frac{1}{(v-z_1)\cdots (v-z_N)} \;=\;
\sum_{i=1}^N
  \prod_{j:\neq i}
   \frac{1}{z_i -z_j}\;
    \frac{1}{v-z_i}\,,
\end{equation}
\hfill $\Box$

\begin{lemma}
\label{highcorr05.lem-Jn2}
The following formula holds
\begin{equation}
\label{highcorr05.eq-Jn2}
J_{\nU}(g;\zU) \;=\;
\sum_{i=1}^N
\sum_{\mU ; |\mU|=n_i}
  \prod_{j:\neq i}
   \frac{(-1)^{m_j} (m_j+n_j)!}{m_j! n_j!}
    \frac{1}{(z_i-z_j)^{m_j+n_j+1}}\;\;
     \frac{1}{m_i!}\; I_0(g^{(m_i)};z_i)\,.
\end{equation}
\end{lemma}

\noindent
{\bf Proof: } Applying (\ref{highcorr05.eq-Jn1}) to
(\ref{highcorr05.eq-Jn11}) gives

$$J_{\nU}(g;\zU) \;=\;
   \sum_{i=1}^N
    \frac{1}{n_i!} \frac{\partial^{n_i}}{\partial z_i^{n_i}}
  \left\{
     \prod_{j:\neq i}
      \frac{1}{(z_i-z_j)^{n_j+1}}\;\;
       I_0(g;z_i)
  \right\}\,.
$$

\noindent The following multiple variable generalization of the Leibnitz
formula will be used

\begin{equation}
\label{highcorr05.eq-leibniz}
\frac{1}{r!}\; \frac{d^r}{dz^r} \;
f_1(z) \cdots f_N(z) \;=\;
  \sum_{\mU; |\mU|=r}
   \prod_{k=1}^N
    \frac{1}{m_k!}\; \frac{d^{m_k}f_k}{dz^{m_k}}
\end{equation}

\noindent Applying it to the previous formula, together with
(\ref{highcorr05.eq-der1survn}), leads to the result.
\hfill $\Box$

\vspace{.2cm}

\noindent The analyticity properties of these functions and their boundary
values is now investigated in a manner similar to Lemma
\ref{highcorr05.lem-Izero}.
For $\sigU \in \{+,-\}^N$, let $J_{\nU}^{\sigU}(g;\EU)$ be defined by

$$J_{\nU}^{\sigU}(g;\EU)\;=\;
   \lim_{\epsilon_k\downarrow 0}
    J_{\nU}(g;E_1+\imath \sigma_1 \epsilon_1, \cdots ,
      E_N+\imath \sigma_N \epsilon_N)\,.
$$

\noindent Then, as a consequence of (\ref{highcorr05.eq-Jn2})

\begin{lemma}
\label{highcorr05.lem-derJn}
If $\sigU \in \{+,-\}^N$,  the function $\EU \in \RM^N \mapsto
J_{\nU}^{\sigU} (g;\EU)\in \CM$ is analytic away from coincident points.
Moreover, away from coincident points,

\begin{eqnarray*}
\frac{1}{\lU!} \;\partial^{\lU} J_{\nU}^{\sigU}(g; \EU) & = &
\sum_{i=1}^N
\sum_{\mU ; |\mU|=l_i+n_i}
  ~\left\{ \prod_{j:\neq i}
   \frac{(-1)^{m_j} (m_j+l_j+n_j)!}{m_j! l_j!n_j!}
    \frac{1}{(E_i-E_j)^{m_j+l_j+n_j+1}} \right\} \\
& & \times
     \frac{(n_i + l_i ) !}{l_i ! n_i ! m_i!}
      \;\; I_0^{\sigma_i}(g^{(m_i)};E_i)\,,
\end{eqnarray*}

\noindent where

$$
I_0^{\sigma_i}(h;E_i) = \lim_{\epsilon_i \downarrow 0}
I_0 ( h; E_i + i  \sigma_i \epsilon_i ) .
$$
\end{lemma}

\noindent
{\bf Proof: } Using equations (\ref{highcorr05.eq-Jn1}) and
(\ref{highcorr05.eq-Jn2}) leads directly to this formula. The real
analyticity comes from the use of the contour of integration used in the
proof of Lemma~\ref{highcorr05.lem-Izero} provided each semicircle has
radius satisfying $0 < \epsilon_i < r$.
\hfill $\Box$

\vspace{.2cm}

\noindent  Let $\Delta_N$ be the usual $(N-1)$-simplex, namely, the set of
$\sU = (s_1, \cdots, s_N) \in [0,1]^N$ such that $|\sU|= s_1 + \cdots + s_N
=1$. Let $d^{N-1}\sU$ denote the measure $ds_1 \cdots ds_{N-1}$ defined on
$\Delta_N$, and set $\sU^{\nU} = \prod_{k=1}^N s_k^{n_k}$. The following 
result generalizes Lemma~\ref{highcorr05.lem-Izero} and 
eq.~\ref{highcorr05.eq-bv1}

\begin{proposi}
\label{highcorr05.prop-JNsigma}
The function $\EU \in \RM^N \mapsto J_{\nU}^{\sigU}(g;\EU)\in \CM$
is analytic away from coincident points and

\begin{equation}
\label{highcorr05.eq-regplussing}
J_{\nU}^{\sigU}(g;\EU)\;=\;
   J_{\nU}^{reg}(g;\EU)+
    \imath \pi \sum_{k=1}^N
     \sum_{\mU,|\mU|=n_k}
      \frac{\sigma_k g^{(m_k)}(E_k)}{m_k!}
       \prod_{j:\neq k}
        \frac{1}{(E_k-E_j)^{n_j+m_j+1}}\,,
\end{equation}

\noindent where $\EU \in \RM^N \mapsto J=J_{\nU}^{reg}(g;\EU)\in \CM$
is analytic everywhere and given by

\begin{equation}
\label{highcorr05.eq-Jnreg}
J=
   \int_{\Delta_N} d^{N-1}\sU
    \left( \frac{\sU^{\nU}}{\nU!} \right)
     \int_0^{\infty} du\;
      \frac{
  g^{(N+|\nU|-1)}(\sum_k s_k E_k +u)-g^{(N+|\nU|-1)}(\sum_k s_k E_k -u)
            }{u}\,.
\end{equation}

\noindent In addition, if $\sigma_1=\cdots =\sigma_N = \pm 1$, the function
$\EU \in \RM^N \mapsto J_{\nU}^{\sigU}(g ; \EU)\in \CM$ is analytic
everywhere and given by

\begin{equation}
\label{highcorr05.eq-sigma=1}
\sigma_1=\cdots =\sigma_N = \pm 1 \qquad\qquad
   \Rightarrow \qquad\qquad
    J_{\nU}^{\sigU}(g ; \EU)\;=\;
     J_{\nU}^{reg}(g ; \EU) \pm\imath \pi R_{\nU}(g ; \EU)\,,
\end{equation}

\noindent with

\begin{equation}
\label{highcorr05.eq-regresidue}
R_{\nU}(g; \EU) \;=\;
   \int_{\Delta_N} d^{N-1}\sU\;
    \left(  \frac{\sU^{\nU}}{\nU!} \right) \; g^{(N+|\nU|-1)}
     \left(\sum_k s_k E_k \right)\,.
\end{equation}
\hfill $\Box$
\end{proposi}

\begin{rem}
\label{highcorr05.rem-singularityofJn}
{ The previous proposition shows that $J_{\nU}^{\sigU}$
admits a polar singularity at
$E_k=E_j$ if and only if $\sigma_k\neq\sigma_j$.}
\end{rem}

\noindent The first step in the proof is a simple lemma relating integration
over an $N-1$-simplex to the product of the singular terms in the integrand
in (\ref{highcorr05.eq-Jn}). Given a set of points $\{ z_1, \cdots , z_N\}$,
their convex hull will be denoted by $\conv{\{z_1, \cdots , z_N\}}$.

\begin{lemma}
\label{highcorr05.lem-conv}
Let $\nU\in\NM^N$ and $\zU\in (\CM\setminus \RM)^N$.
If $v\notin \conv{\{z_1, \cdots , z_N\}}$, the following formula holds

$$\prod_{k=1}^N
   \frac{1}{(v-z_k)^{n_k+1}} \;=\;
    \frac{(N+|\nU|-1)!}{\nU !}\;
     \int_{\Delta_N} d^{N-1}\sU \;
      \frac{\sU^{\nU}}{(v-\sum_{k=1}^N s_k z_k)^{N+|\nU|}}\,.
$$
\end{lemma}

\noindent
{\bf Proof: } Both sides of this formula are defined and holomorphic in $v$
in the domain $\{ v\in \CM\,;\, v\notin \conv\{z_1, \cdots, z_N\}\}$. Thus,
using the unique analytic continuation theorem, it is sufficient to prove it
for $\Re{(v-z_k)} >0$. Setting $a_k= v-z_k$, the identity

$$\frac{1}{a_k^{n_k+1}} \;=\;
   \int_0^{\infty} dt_k \;\frac{t_k^{n_k}}{n_k!} \; e^{-t_k a_k},
$$

\noindent valid for $\Re{a_k >0}$, implies

$$\prod_{k=1}^N
   \frac{1}{(a_k)^{n_k+1}} \;=\;
    \int_{\RM_+^N} d^N\tU\;
     \frac{\tU^{\nU}}{\nU!}\; e^{-\sum_k t_ka_k}\,.
$$

\noindent The following change of variables $(t_1, \cdots, t_N) \mapsto
(\lambda, s_1, \cdots, s_{N-1})$ will be useful

$$t_k \;=\; \lambda s_k,
\hspace{1cm} (1\leq k\leq N) \hspace{1cm}
s_N = 1- \sum_{k=1}^{N-1} s_k\,,\; \lambda \geq 0\,.
$$

\noindent In particular, we have $s_k \geq 0\; \forall k$, and $s_1 + \cdots
+ s_{N-1}\leq 1$. Furthermore, the volume elements transform as

$$dt_1 \wedge \cdots \wedge dt_N \;=\; (-1)^{N-1}
   \lambda^{N-1} d\lambda \wedge ds_1 \wedge \cdots \wedge ds_{N-1}\,,
$$

\noindent so that

\begin{eqnarray*}
\prod_{k=1}^N
   \frac{1}{(a_k)^{n_k+1}} &=&
    \int_{\Delta_N} ds_1\cdots ds_{N-1}\;
     \frac{\sU^{\nU}}{\nU!}\;
     \int_0^{\infty} \lambda^{N+|\nU|-1} \,d\lambda
      e^{-\lambda \sum_k s_ka_k}\\
& = &
   \int_{\Delta_N} ds_1\cdots ds_{N-1}\;
    \frac{\sU^{\nU}}{\nU!}\;
     \frac{(N+|\nU|-1)!}{(\sum_k s_ka_k)^{N+ |\nU|}}\,.
\end{eqnarray*}

\noindent Replacing $a_k$ by $v-z_k$ gives the result.
\hfill $\Box$

\vspace{.2cm}

\noindent
{\bf Proof of Proposition~\ref{highcorr05.prop-JNsigma}: }

\noindent
1. The analyticity claim in the first part of the proposition
follows once we have proved the representations
(\ref{highcorr05.eq-regplussing}) and (\ref{highcorr05.eq-Jnreg}).  If $z_1,
\cdots, z_N$ are all on the same side of the real axis, then the convex hull
is also contained in the same half plane.
Thanks to Lemma~\ref{highcorr05.lem-conv}, it follows that

\begin{eqnarray*}
J_{n_1,\cdots,n_N}(g;z_1, \cdots, z_N) & = &
   \int_{\Delta_N} d^{N-1}\sU\;
    \frac{\sU^{\nU}}{\nU!}\;
     \int_{-\infty}^{+\infty}
      dv\;
      \frac{g(v)(N+|\nU|-1)!}{(v-\sum_k s_k z_k)^{N+|\nU|}}\\
& = &
  \int_{\Delta_N} d^{N-1}\sU\;
   \frac{\sU^{\nU}}{\nU!}\;
    I_0 \left(
     g^{(N+|\nU|-1)}\,;\, \sum_k s_k z_k
        \right)
\end{eqnarray*}

\noindent From Lemma~\ref{highcorr05.lem-Izero}, equations
(\ref{highcorr05.eq-Jnreg}), (\ref{highcorr05.eq-sigma=1}), and
(\ref{highcorr05.eq-regresidue}) follow immediately.

\vspace{.1cm}
\noindent
2. To derive (\ref{highcorr05.eq-regplussing}) from
(\ref{highcorr05.eq-regresidue}) in the case $\sigma_1 = \cdots = \sigma_N =
\pm 1$, we apply Lemma~\ref{highcorr05.lem-conv} for $n_1=\cdots=n_N=1$,
together with (\ref{highcorr05.eq-primefactor}), to obtain

$$\int_{\Delta_N} d^{N-1}\sU\;
   \frac{(N-1)!}{(z-\sum_k s_k E_k)^N} \;=\;
    \frac{1}{(z-E_1)\cdots(z-E_N)}\;=\;
     \sum_{k=1}^N
      \frac{1}{z-E_k}
       \prod_{j:\neq k}
        \frac{1}{E_k-E_j}\,.
$$

\noindent Multiplying both sides by $g(z)$ and integrating over a Jordan
path surrounding each $E_k$ once and contained in the holomorphy domain of
$g$, leads to

$$\int_{\Delta_N} d^{N-1}\sU\;
   g^{(N-1)}\left(\sum_k s_k E_k\right)
    \;=\;
     \sum_{k=1}^N
      g(E_k)
       \prod_{j:\neq k}
        \frac{1}{E_k-E_j}\,.
$$

\noindent Applying the differential operator $(1/\nU!) ~
\partial_{\EU}^{\nU}$ on both sides gives

$$\int_{\Delta_N} d^{N-1}\sU\;
   \frac{\sU^{\nU}}{\nU!}\;
   g^{(N+|\nU|-1)}\left(\sum_k s_k E_k\right)
    \;=\;
   \sum_{k=1}^N
    \frac{1}{n_k} \partial_{E_k}^{n_k}
     \left(
      g(E_k)
       \prod_{j:\neq k}
        \frac{1}{(E_k-E_j)^{n_j+1}}
     \right)\,.
$$

\noindent An application of Leibnitz rule (see 
(\ref{highcorr05.eq-leibniz}))
gives equation (\ref{highcorr05.eq-regplussing}) for $\sigma_1=\cdots
=\sigma_N =1$.

\vspace{.1cm}
\noindent
3. If $\sigU=(\sigma_1,\cdots,\sigma_N)\in \{+1,-1\}^N$ is arbitrary,
let $A(\sigU)\subset [1,N]$ be the set of indices $k$ such that
$\sigma_k =-1$. Whenever $\zU = (z_1, \cdots, z_N)\in \CM^N$
is such that $\sigma_k \Im{z_k} >0$ for all $k$'s,
the integration contour defining $J_{\nU}(g;\zU)$
cannot be deformed to have all $z_k$'s on the same side.
Actually the $z_k$'s such that $k\in A(\sigU)$ are below the
integration path while the other $z_k$'s are above.
Deforming the contour to get all the $z_k$'s above
can be done to the price of adding the corresponding residues,
leading to

\begin{eqnarray*}
J_{\nU}^{\sigU}(\EU) &=&
   J_{\nU}^{+}(\EU) - 2\imath\pi
    \sum_{k\in A(\sigU)}
     \Res_{z=E_k}
      \frac{g(z)}{(z-E_1)^{n_1+1}\cdots (z-E_N)^{n_N+1}}\\
&=& J_{\nU}^{+}(\EU) + \imath\pi
    \sum_{k=0}^N (\sigma_k -1)
     \Res_{z=E_k}
      \frac{g(z)}{(z-E_1)^{n_1+1}\cdots (z-E_N)^{n_N+1}}\,.
\end{eqnarray*}

\noindent The residues are given by

$$\Res_{z=E_k}
      \frac{g(z)}{(z-E_1)^{n_1+1}\cdots (z-E_N)^{n_N+1}}\;=\;
  \frac{1}{n_k!} \partial_{E_k}^{n_k}
   \left(
    g(E_k) \prod_{j:\neq k} \frac{1}{(E_k-E_j)^{n_j+1}}
   \right)
$$

\noindent
Use of the Leibnitz rule again together with the previous formula for
$J_{\nU}^{+}(g;\EU)$ gives (\ref{highcorr05.eq-regplussing})
in the general case.
\hfill $\Box$

\vspace{.3cm}

\section{Real Analyticity}
\label{highcorr05.sect-realanal}

\noindent We now use the estimates on the Cauchy integrals obtained
in Section \ref{highcorr05.sect-Ncauchy} in order to prove that the
$N$-point Cauchy-type integrals, for $N \geq 2$, are real analytic as
functions of $\underline{E} \in \NM^N$, on the domain for which $E_j \neq
E_i$, for $i \neq j$.
To this end, we prove that the functions
$J_{\underline{n}}^{\underline{\sigma}} (g; \underline{E})$, defined in
section 6, have uniformly convergent power series on appropriate domains. It
is convenient to do this using certain Banach spaces of
real analytic functions.
Let $\delta >0$ and let $\Uu$ be an open subset of $\RM^N$.
For a continuous complex valued function $f$ on $\Uu$ let $\|f\|_{\Uu}$
be the sup-norm  $\|f\|_{\Uu} = \sup_{\EU\in \Uu}|f(\EU)|$.
Then let $\Ee_{\delta}(\Uu)$ be the space of
smooth functions $F:\EU \in \Uu \mapsto \CM$ vanishing at infinity, together
with all derivatives, such that

$$\|F\|_{\delta,\, \Uu}\;=\;
   \sum_{\lU\in \NM^N}
    \frac{\delta^{|\lU|}}{\lU!}\;
     \|\partial^{\lU} F \|_{\Uu} \; <\; \infty\,.
$$

\noindent Endowed with the norm $\|\cdot \|_{\delta,\, \Uu}$ the space
$\Ee_{\delta}(\Uu)$ is a Banach space. Moreover, it is a Banach
$\ast$-algebra if endowed with the pointwise multiplication and complex
conjugacy since

\begin{lemma}
\label{highcorr05.lem-Eealg}
Let $F$ and $G$ be two elements of $\Ee_{\delta}(\Uu)$.
Then both $F^{\ast}$ and $F\cdot G$ belong to $\Ee_{\delta}(\Uu)$ and
$$\|F\cdot G\|_{\delta,\, \Uu} \;\leq \;
   \|F\|_{\delta,\, \Uu} \; \| G\|_{\delta,\, \Uu}\,,
\hspace{2cm}
  \|F^{\ast}\|_{\delta,\, \Uu} \;=\; \|F\|_{\delta,\, \Uu}
$$
\end{lemma}

\noindent
{\bf Proof: } The only non trivial property is the first inequality. Using
the Leibnitz rule (see (\ref{highcorr05.eq-leibniz})) leads to

$$\frac{1}{\lU!}\;\partial^{\lU} \left( F\cdot G\right) \;=\;
   \sum_{\mU,\nU; \mU+\nU=\lU}
    \frac{1}{\mU!}\;\partial^{\mU} F\;
     \frac{1}{\nU!}\;\partial^{\nU} G\,.
$$

\noindent From this the inequality follows
immediately since $\delta^{|\mU+\nU|} = \delta^{|\mU|}\, \delta^{|\nU|} $.
\hfill $\Box$

\vspace{.1cm}

\noindent
When we take $\Uu = \RM^N$, we write $\| \cdot \|_\delta$ in place of $\|
\cdot \|_{\delta, \RM^N}$.

\begin{proposi}
\label{highcorr05.prop-Jnanalytic}
Let $g$ be an element of $\Hh_r$, for some $r>0$, let $C=8/\pi+ 2+r+r^2$,
and consider $N \geq 2$. For any choice of $\delta > 0$ and $\Delta > 0$
such that $0<\delta < \Delta/2$ and $0 < \Delta -\delta < r/2$, the boundary
values of $J_{\nU}^{\sigU}(g)$ satisfy the following estimate

$$\|J_{\nU}^{\sigU}(g)\|_{\delta,\, \Dd_{\Delta}} \; \leq \;
   \frac{4CN\|g\|_r}{r} \left( \frac{e}{\Delta -\delta} \right)^{|\nU|+N}\,.
$$

\noindent in the domain $\Dd_{\Delta}= \{\EU\in \RM^N\,;\,
|E_i-E_j|>\Delta\}$.
\end{proposi}

\noindent The proof will proceed in several steps. First the following
identity will be used often

\begin{equation}
\label{highcorr05.eq-taylor}
(1-u)^{-n-1}\;=\;
   \sum_{l=0}^{\infty}
    \frac{(l+n)!}{l!\;n!} \;u^l\,.
\end{equation}

\noindent The following combinatorial estimate is necessary.

\begin{lemma}
\label{highcorr05.lem-combestim}
If $\nU\in\NM^L$ and if $r \in \NM$, then

\begin{equation}
\label{highcorr05.eq-combid2}
\sum_{|\mU|=r}
   \frac{(\mU+\nU)!}{\mU!\;\nU!} \;=\;
    \frac{(r+|\nU|+L-1)!}{r!\;(|\nU|+L-1)!}
\end{equation}
\end{lemma}

\noindent  {\bf Proof: } If $C_r^L(\nU)$ denotes the
left side of (\ref{highcorr05.eq-combid2}),
its definition and the identity (\ref{highcorr05.eq-taylor}) give

$$\sum_{r=0}^{\infty}
   C_r^L(\nU) X^r \;=\;
   \prod_{k=1}^L
    \sum_{m_k=0}^{\infty}
     \frac{(m_k+n_k)!}{m_k!\;n_k!}\; X^{m_k}\;=\;
    \frac{1}{(1-X)^{|\nU|+L}} .
$$

\noindent Expanding the right side
in formal power series in $X$ by using
(\ref{highcorr05.eq-taylor}) again gives directly the result.
\hfill $\Box$

\vspace{.1cm}

\noindent {\bf Proof of Proposition~\ref{highcorr05.prop-Jnanalytic}: }
Using Lemma~\ref{highcorr05.lem-derJn} and assuming $|E_i-E_j|>\Delta$ gives

\begin{eqnarray*}
\|J_{\nU}^{\sigU}(g)\|_{\delta, \Dd_{\Delta}} &\leq &
\sum_{i=1}^N
  \sum_{\lU\in\NM^N}
   \sum_{\mU ; |\mU|=l_i+n_i}
    ~\left\{ \prod_{j:\neq i}
   \frac{(m_j+l_j+n_j)!}{m_j!\; l_j!\; n_j!}
    \left(
        \frac{\delta}{\Delta}
    \right)^{l_j} \frac{1}{\Delta^{m_j+n_j+1}} \right\} \\
& & \times
     \frac{(l_i+n_i)!}{l_i!\; n_i!}\; \delta^{l_i}\;
      \|I_0\left( \frac{g^{(m_i)}}{m_i!}\right)\|_{\infty}\,.
\end{eqnarray*}

\noindent Using the distributive property of the product with respect to
addition for the variable $l_j$ gives

\begin{eqnarray}
\label{highcorr05.eq-expansion11}
\|J_{\nU}^{\sigU}(g)\|_{\delta, \Dd_{\Delta}} &\leq &
\sum_{i=1}^N
  \sum_{l_i=0}^{\infty}
   \sum_{\mU ; |\mU|=l_i+n_i}
    ~\left\{ \prod_{j:\neq i}
     ~\left( \sum_{l_j=0}^{\infty}
   \frac{(m_j+l_j+n_j)!}{m_j!\; l_j!\; n_j!}
    \left(
        \frac{\delta}{\Delta}
    \right)^{l_j} \right) \frac{1}{\Delta^{m_j+n_j+1}} \right\} \nonumber \\
& & \times
     \frac{(l_i+n_i)!}{l_i!\; n_i!}\; \delta^{l_i}\;
      \|I_0\left( \frac{g^{(m_i)}}{m_i!}\right)\|_{\infty}\,.
\end{eqnarray}

\noindent We use Lemmas \ref{highcorr05.lem-derivative} and
\ref{highcorr05.lem-Izeroestim} to estimate the $L^\infty$-norm in
(\ref{highcorr05.eq-expansion11}) for any $0 < r_i < r$ by

\begin{equation}
\label{highcorr05.eq-expansion12}
\|I_0 \left( \frac{g^{(m_i)}}{m_i!}\right)\|_\infty
\leq
  \frac{C}{(r-r_i)^2}
   \frac{ \| g^{(m_i)} \|_{r-r_i}}{m_i !}\,,
\hspace{2cm}
C=\frac{8}{\pi}+ 2+ r+r^2\,.
\end{equation}

\noindent
Using (\ref{highcorr05.eq-taylor}) to sum over $l_j$, together with estimate
(\ref{highcorr05.eq-expansion12}), leads to

\begin{eqnarray*}
\|J_{\nU}^{\sigU}(g)\|_{\delta, \Dd_{\Delta}} &\leq &
\sum_{i=1}^N
  \sum_{l_i=0}^{\infty}
   \sum_{\mU ; |\mU|=l_i+n_i}
    ~\left\{ \prod_{j:\neq i}
     \frac{(m_j+n_j)!}{m_j!\; n_j!}
      \frac{1}{(\Delta-\delta)^{m_j+n_j+1}} \right\} \\
& & \times
     \frac{(l_i+n_i)!}{l_i!\; n_i!}
      \frac{\delta^{l_i}}{r_i^{m_i}}\;
       \frac{C}{(r-r_i)^2} \|g\|_r\,,
\end{eqnarray*}

\noindent for $0<r_i<r$. We define $p$ by $|\mU| = m_i+p$ and $|\nU|_i=
|\nU|-n_i$. With this definition and Lemma~\ref{highcorr05.lem-combestim},
we perform the restricted sum over $\underline{m} \in \NM$ and obtain

\begin{eqnarray*}
\|J_{\nU}^{\sigU}(g)\|_{\delta, \Dd_{\Delta}} &\leq &
\sum_{i=1}^N
    \frac{1}{(\Delta-\delta)^{|\nU|_i+N-1}}
  \sum_{l_i=0}^{\infty}
   ~\left(  \sum_{p=0}^{l_i+n_i}
     \frac{(p+|\nU|_i+N-2)!}{p! (|\nU|_i+N-2)!}
      \left(
      \frac{r_i}{\Delta-\delta}
      \right)^p \right) \nonumber  \\
& & \times
     \frac{(l_i+n_i)!}{l_i!\; n_i!}
      \frac{\delta^{l_i}}{r_i^{\,n_i+l_i}}\;
       \frac{C}{(r-r_i)^2} \|g\|_r\,.
\end{eqnarray*}

\noindent Using (\ref{highcorr05.eq-taylor}) again to sum over $p$, if $r_i
<\Delta -\delta$, leads to

\begin{eqnarray*}
\|J_{\nU}^{\sigU}(g)\|_{\delta, \Dd_{\Delta}} &\leq &
\sum_{i=1}^N
  \frac{C \|g\|_r}{(r-r_i)^2r_i^{\,n_i}(\Delta -\delta-r_i)^{|\nU|_i+N-1}}
  \sum_{l_i=0}^{\infty}
     \frac{(l_i+n_i)!}{l_i!\; n_i!}
      \left(
       \frac{\delta}{r_i}
      \right)^{l_i}\;
       \,.
\end{eqnarray*}

\noindent Finally, using again (\ref{highcorr05.eq-taylor}) to sum over
$l_i$,
if $\delta <r_i <r$, leads to

\begin{eqnarray*}
\|J_{\nU}^{\sigU}(g)\|_{\delta, \Dd_{\Delta}} &\leq &
\sum_{i=1}^N
  \frac{C r \|g\|_r}{(r-r_i)^2(r_i-\delta)^{\,n_i+1}
      (\Delta -\delta-r_i)^{|\nU|_i+N-1}}\,.
\end{eqnarray*}

\noindent These estimates are satisfied on the domain $\Dd_{\Delta}$
provided $0<\delta < r_i < r$, and $(\Delta-\delta) > 0$. In particular, if
we take $\Delta >2\delta$. Recalling that we treat here $N\geq 2$, if
$\Delta -\delta <r/2$,
it follows that $r_i<r/2$, so that $r/(r-r_i)^2 <4/r$. Then, minimizing over
$r_i$ in each term on the right side gives

$$r_i \;=\;
   \frac{(n_i+1)(\Delta -\delta) + (|\nU|_i+N-1)\delta}{|\nU|+N}\,,
$$
\noindent leading to

$$r_i-\delta = (\Delta-\delta)\frac{n_i+1}{|\nU|+N}\,,
  \hspace{2cm}
   \Delta -\delta -r_i = (\Delta-\delta)\frac{|\nU|_i+N-1}{|\nU|+N}\,.
$$

\noindent If $l,m$ are positive integers,
$((l+m)/l)^l =(1+m/l)^l\leq e^m$, so that

$$\left( \frac{l+m}{l}\right)^l\;
   \left(  \frac{l+m}{m}\right)^m \;\leq\;
     e^{l+m}\,,
$$

\noindent we obtain

$$\|J_{\nU}^{\sigU}(g)\|_{\delta, \Dd_{\Delta}} \;\leq \;
   \frac{4NC\|g\|_r}{r}\;
    \left(\frac{e}{\Delta-\delta}\right)^{|\nU|+N}\,,
\hspace{1cm}
   \mbox{\rm if  }\;\; 0< \delta<\Delta-\delta < \frac{r}{2}\,.
$$
This proves Proposition 3.
\hfill $\Box$

\vspace{.3cm}

  \section{Proof of Theorem~\ref{highcorr05.th-smoothG}}
  \label{highcorr05.sect-pfmaintheorem}

\noindent It is now possible to finish the proof of
Theorem~\ref{highcorr05.th-smoothG} proving the main theorem on the real
analyticity of the $N$-point correlation functions $G_{\AU} (\EU )$, away
from a small neighborhood of the coincident points
$E_i =  E_j$. We recall the $\AU$-compatible path expansion in
(\ref{highcorr05.eq-Nptrwe}).
Our strategy is to write $G_{\AU} (\EU )$ as a sum over $n \in \NM$, the
length of
the compatible paths. We then prove that $G_{\AU} (\EU )$ is in the Banach
space
$\Ee_{\delta}(\Dd_{\Delta})$ using the estimates in Proposition 3.
We first prove the theorem
for the case when the covariant operators $A_i$ are $r$-monomials
which implies the result when the covariant operators $A_i$ are
$r$-polynomials.
Starting from (\ref{highcorr05.eq-Nptrwe}), it is sufficient to
prove that the formal path expansion converges
in $\Ee_{\delta}(\Dd_{\Delta})$, with the domain $\Dd_{\Delta} \subset
\RM^N$,
for convenient values of $\delta$ and
$\Delta$. Using Proposition~\ref{highcorr05.prop-Jnanalytic},
the $N$-point correlation $G_{\AU}^{\sigU}$ is estimated as follows

$$\|G_{\AU}^{\sigU}\|_{\delta,\Dd_{\Delta}} \;\leq\;
   \sum_{\Gamma\in \ps(\AU)}
    |\lambda|^{|\Gamma|}
     \prod_{u\in\Vv(\Gamma)}
      NC_1\|g_{\Gamma, u}\|_r \;
       \left(\frac{e}{\Delta-\delta}\right)^{|\nU_{\Gamma}(u)-\fU|+N}\,,
\hspace{1cm}
C_1= \frac{4C}{r}\,.
$$

\noindent Clearly, we have that $|\nU_{\Gamma}(u)-\fU|+N =
|\nU_{\Gamma}(u)|$. This is bounded above by the length of the path $|
\Gamma|$. On the other hand,
if $A=\max_{i,u} \|a_{i,u}\|$, then $A$ appears at most $N$-times in the
right hand side in the definition of $g_{\Gamma, u}$, reflecting the fact
that $G_{\AU}$
is homogeneous of degree $N$ in $\AU$. As a consequence, we get the bound

$$\prod_{u\in\Vv(\Gamma)}
      NC_1 \|g_{\Gamma, u}\|_r \;\leq\;
   A^N \prod_{u\in\Vv(\Gamma)}
      NC_1 \|g\|_r.
$$

\noindent On the other hand, since $g$ defines a probability on $\RM$,
$\|g\|_r \geq 1$. Moreover, $C_1= 4(8/\pi+ 2)/r+ 4+ 4r\geq 4$, so that $NC_1
\|g\|_r >1$. Hence, using $\#\Gamma \leq |\Gamma|$, and the
Lemma~\ref{highcorr05.lem-countingpath}, this inequality becomes

\begin{eqnarray}\label{mainestG1}
\|G_{\AU}^{\sigU}\|_{\delta,\Dd_{\Delta}} & \leq & A^N
   \sum_{\Gamma\in \ps(\AU)}
    \left(\frac{e|\lambda|}{\Delta-\delta}\right)^{|\Gamma|}
     \prod_{u\in\Vv(\Gamma)}
      NC_1\|g\|_r \;\leq\; A^N
       \sum_{\Gamma\in \ps(\AU)}
    \left(\frac{e|\lambda|NC_1\|g\|_r}{\Delta-\delta}\right)^{|\Gamma|} 
\nonumber \\
  & \leq &  A^N \sum_{n=0}^{\infty} 
\left(\frac{2dNC_1e\,|\lambda|\|g\|_r}{\Delta-\delta}\right)^n
            \;\leq \;
      \frac{A^N}{1- 2dNC_1e|\lambda|\, \|g\|_r (\Delta-\delta)^{-1}} <  
\infty  ,  \nonumber \\
\end{eqnarray}

\noindent provided

$$
|\lambda |\;<\;   \frac{\Delta-\delta}{2dNC_1e\,\|g\|_r}.
$$

\noindent Choosing $\delta < \Delta / 2$, we define $a_0 \equiv 4 d N C_1 e
\|g \|_r$. Hence,
the function $G_{\AU}^{\sigU} \in \Ee_{\delta}(\Dd_{\Delta})$, provided
$a_0 | \lambda | < \Delta$, proving Theorem 1 for the case when 
$\underline{A}$ is
a covariant family of $r$-monomials. We now consider a general covariant 
family $\underline{A} \in
L^\infty ( \mathcal{T}_\PM )^{\times N}$ and the associated
correlation function $G_{\underline{A}} (\underline{z})$,
with $\Im z_j > 0$. By Proposition 1, $G_{\underline{A}} (\underline{z})$ 
can be approximated
uniformly on any compact subset of $(\CM \backslash \RM )^N$ by a sequence 
of correlation functions
$G_{\underline{A}_n} (\underline{z})$, where the family $\underline{A}_n$ 
are $r$-polynomials.
Furthermore, the coefficients can be chosen so that they are uniformly
bounded in $L^\infty ( \Omega, \PM)$.
This family of correlation functions has an analytic
continuation to $\Dd_{\Delta}$. Estimate (\ref{mainestG1})
show that this family of approximating correlations functions is
uniformly bounded since the bound depends on the coefficients.
Since the family converges uniformly on any compact subset of the original
domain $( \CM \backslash \RM)^N$, the family
converges uniformly on any compact subset of the extended domain in $\CM^N$.
The limit functions provide a continuation of $G_{\underline{A}} 
(\underline{z})$ into this
extended domain by the identity principle for analytic functions.
This proves Theorem 1.
\hfill $\Box$
\vspace{.3cm}

%
%

\end{document}